# Engineering Relaxation Pathways in Building Blocks of Artificial Spin Ice for Computation

H. Arava[a,b,*], N.R. Leo[a,b,†,*], D. Schildknecht[a,b,c], J. Cui[a,b], J. Vijayakumar[d],
P.M. Derlet[c], A. Kleibert[d], L.J. Heyderman[a,b]

[a]Laboratory for Mesoscopic Systems, Department of Materials, ETH Zurich, 8093 Zurich, Switzerland
[b]Laboratory for Multiscale Materials Experiments, Paul Scherrer Institute, 5232 Villigen PSI, Switzerland
[c]Condensed Matter Theory Group, Paul Scherrer Institute, 5232 Villigen PSI, Switzerland
[d]Swiss Light Source, Paul Scherrer Institute, 5232 Villigen PSI, Switzerland

Nanomagnetic logic, which makes use of arrays of dipolar-coupled single domain nanomagnets for computation, holds promise as a low power alternative to traditional computation with CMOS. Beyond the use of nanomagnets for Boolean logic, nanomagnets can also be exploited for non-deterministic computational schemes such as edge detection in images and for solving the traveling salesman problem. Here, we demonstrate the potential of arrangements of thermally-active nanomagnets based on artificial spin ice for both deterministic and probabilistic computation. This is achieved by engineering structures that follow particular thermal relaxation pathway consisting of a sequence of reorientations of magnet moments from an initial field-set state to a final low energy output state. Additionally, we demonstrate that it is possible to tune the probability of attaining a particular final low-energy state, and therefore the likelihood of a given output, by modifying the intermagnet distance. Finally, we experimentally demonstrate a scheme to connect several computational building blocks for complex computation.

*Email: hanu.arava@psi.ch; n.leo@nanogune.eu
[†]Currently at: cCIC nanoGUNE, 20018 Donostia-San Sebastian, Spain

## I. INTRODUCTION

All-magnetic computation with arrays of dipolar-coupled nanomagnets has the potential to provide a low-power alternative to existing CMOS technologies [1-6]. In the last decade, significant progress has been made in the use of nanomagnets for deterministic computation, which is often referred to as nanomagnetic logic, implementing nanomagnets with in-plane [1,2,7] and perpendicular anisotropy [8] to perform Boolean operations. Here, it is the two stable magnetic configurations associated with each magnet that can be used to encode binary information.

A variety of approaches have been developed to apply nanomagnets to computation, including strain-mediated logic [9], spin-Hall-induced effects [10], as well as their three-dimensional implementation [11]. In addition, a high clocking speed in a chain of dipolar coupled nanomagnets was achieved using a pulsed current to transfer information at ultrafast timescales [12]. Beyond Boolean computation, novel computational schemes with nanomagnets have emerged such as designing arrays of nanomagnets where the relaxation to a low energy state after the application of a field can be implemented for image recognition [13] or using coupled randomly switching nanomagnets to solve, for example, the traveling salesman problem [14].

The exploitation of more complex arrangements of nanomagnets such as those found in artificial spin ice, where nanomagnets are placed on the sites of a periodic lattice such as a square [15] or kagome lattice [16], brings advantages such as an improved reliability of computation, which we have demonstrated in structures based on artificial square ice [17]. In order to exploit artificial spin ice structures for applications, it is useful to be able to access specific moment configurations. This has been achieved, for example, in small artificial kagome spin ice structures consisting of up to three hexagonal rings of nanomagnets with specific field protocols or by modifying the shape of individual nanomagnets [16,18]. Other avenues to access specific magnetic states in artificial spin ice involved the use of a specialized magnetic tip [19,20]. Finally, there have been proposals to use artificial kagome spin ice structures [21], as well as to combine multiple three-moment logic gates in tree-like circuits [22] for computation.

In this work, we control the energy landscape associated with artificial spin ice, engineering the relaxation pathways in structures containing arrays of thermally-active dipolar-coupled nanomagnets. These relaxation pathways consist of a sequence of reorientations of the nanomagnet moments as the structures relax from an initial field-set state to a final low-energy state. Based on measurements of the thermal evolution of the magnetic configurations, as well as kinetic Monte-Carlo simulations, we identify two distinct types of pathways in the energy landscape that are either deterministic or probabilistic. For a deterministic relaxation from an initial field-set state, the final state is expected to always be a particular low energy state. In contrast, for probabilistic relaxation, there are two or more possible final low energy states, which occur with a certain probability within a particular time-frame.

Here we present strategies to tune the probability of low-energy states. Finally, we experimentally demonstrate a way to integrate individual computational units into large circuits. With this, we provide a foundation for the implementation of artificial spin ice for deterministic and probabilistic computation.

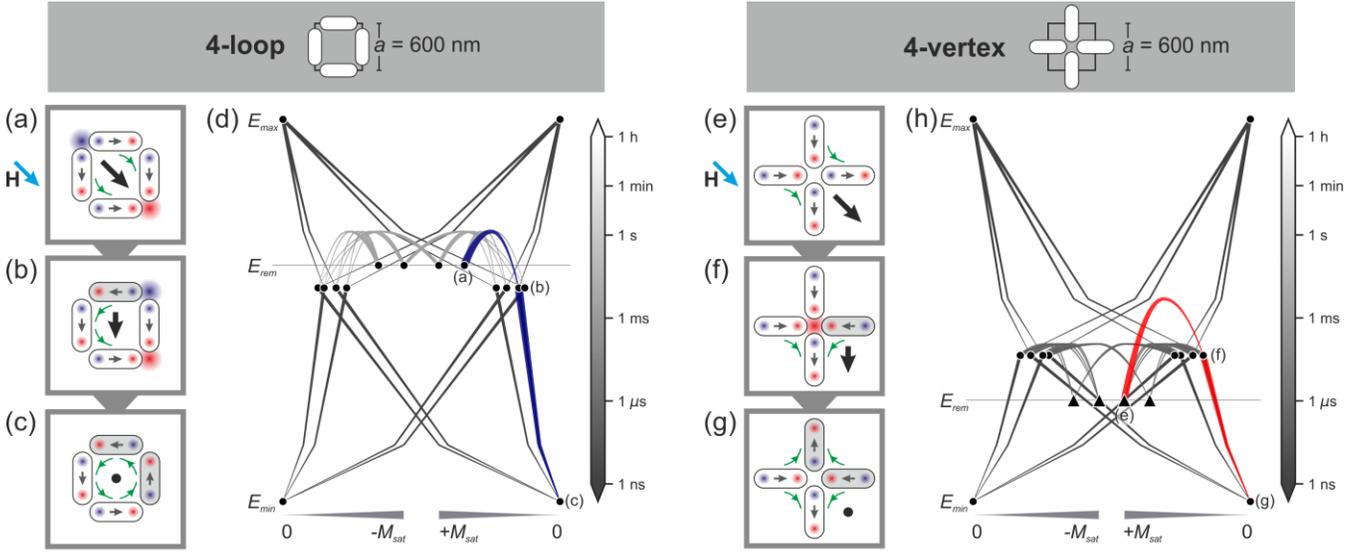

**Fig. 1** Thermal relaxation and state network diagrams for 4-loop and 4-vertex structures. (a-c) Thermal relaxation in a 4-loop structure from a field-set state to one of the two degenerate flux-closure ground states. Switched magnets are highlighted in grey, head-to-tail arrangements are indicated by green arrows and the net magnetization by black arrows. (d) State network diagram for the 4-loop structure. The states are indicated with black points that are located according to their energies along the vertical axis and the net magnetization of the state along the horizontal axis. Transitions between states $m$ and $l$ via single-nanomagnet reorientations are indicated by tapered lines. The midpoint of the lines corresponds to the energy $E_m + E_T$, where the barrier energy $E_T = E_b + \frac{1}{2}(E_l - E_m)$ can be positive or negative. The grayscale of the lines corresponds to the transition rate at $T = 295$ K according to Eq. (2) and the greyscale bar. The blue line represents the transitions between the states shown in (a-c), resulting in a monotonous relaxation path with successive lowering of dipolar energy. (e-g) In a 4-vertex structure, thermal relaxation from a field-set state to the ground state requires a transition to a state of higher dipolar energy, which is associated with the creation of a magnetic charge at the vertex highlighted in red in panel (f). (h) State network diagram for the 4-vertex structure. The red intermittent path is for the state evolution shown in (e-g), where an intermediate state with higher dipolar energy is accessed. The metastable states are indicated with triangles while the ground states are indicated with larger black points.

## II. BACKGROUND

Due to their shape anisotropy, elongated nanomagnets made from soft magnetic materials, such as Permalloy ($Ni_{80}Fe_{20}$), form single-domain states with a net magnetic moment pointing along the long axis of the nanomagnet, and thus can be represented by macrospins with Ising degrees of freedom. For arrangements of dipolar-coupled nanomagnets such as those shown in Fig. 1, the interaction potential, assuming point-like dipole moments, is given by:

$$V(r_{ij}, m_i, m_j) = -\frac{\mu_0}{4\pi r_{ij}^3}[3(m_i \cdot \hat{r}_{ij})(m_j \cdot \hat{r}_{ij}) - m_i \cdot m_j] \quad (1),$$

with the distance vector $r_{ij}$ separating the magnetic point dipoles $m_i$ and $m_j$, and $\hat{r}_{ij} = \frac{r_{ij}}{|r_{ij}|}$. The total energy of an arrangement of $N$ macro-spins is then given by $E_{total} = \frac{1}{2}\sum_{i,j=1}^{N} V(r_{ij}, m_i, m_j)$. Dipolar interactions favor flux-closure head-to-tail configurations, whereas head-to-head or tail-to-tail configurations are energetically unfavorable. Furthermore, the dynamics of thermally-assisted, spontaneous moment reorientations of individual nanomagnets is described by a transition rate, which is given by the moment reorientation rate $v(E_T, T)$, where:

$$v(E_T, T) = v_0 e^{-\frac{E_T}{k_B T}}. \quad (2)$$

Here, $v_0$ is the attempt frequency, $k_B$ is the Boltzmann constant, and $T$ the temperature. The barrier energy $E_T = E_b + \frac{1}{2}\Delta E_{m \to l}$ is the energy required for a single magnet in the array to switch and includes the single-nanomagnet switching barrier energy $E_b$, which is modified by the difference in dipolar energy $\Delta E_{m \to l}$ between the initial and final nanomagnetic states $m$ and $l$ that are separated by a single moment reorientation [23,24]. In the limit of low temperatures, the reorientation dynamics are slow, and thermal relaxation from an initial state towards a state of lower dipolar energy will proceed via successive reversals of individual magnetic moments. In configuration space, the system will therefore follow a path that is defined by a sequence of single-moment reversals. Such relaxation paths have a sequence of moment reorientations that can be classified into two categories: monotonous and intermittent. For a *monotonous* relaxation path, every moment reorientation results in a reduction of dipolar energy, i.e. $E_{i+1} < E_i$ for all states $i$. In contrast, *intermittent* relaxation paths are characterized by the presence of at least one transition to a higher-energy state in its sequence, i.e. $E_{i+1} > E_i$ for at least one $i$. Due to the exponential relationship between the transition rate and the barrier energy in Eq. (2), transitions towards states of higher dipolar energy will be kinetically suppressed.

Whether the paths are monotonous or intermittent strongly depends on the spatial arrangement of the nanomagnets. This is illustrated by the relaxation behavior from an initial field-set configuration to the ground state for two basic motifs with four nanomagnets derived from artificial square ice shown in Fig. 1 [15]. For the 4-loop structure shown in Fig. 1(a-c), the nanomagnets are arranged in a loop, and for the 4-vertex structure shown in Fig 1(e-g), the nanomagnets meet at a common vertex. In both cases, the initial field-set state with energy $E$ is set by applying a magnetic field $H$, indicated by a blue arrow in Figs. 1(a) and (e). On removing the field, the individual nanomagnets will undergo thermally-induced moment reorientations and the nanomagnet structure will relax to a ground state with minimal energy $E_{min}$. For the 4-loop structure, the progression from the initial to the final, low energy flux-closure state [Fig. 1(a-c)], involves the successive lowering of dipolar energy and is thus a monotonous relaxation path. In contrast, the relaxation of the 4-vertex structure to the ground state [Fig. 1(e-g)], requires a transition from a state with one head-to-head and one tail-to-tail interaction [Fig. 1(e)] to a state with three head-to-head arrangements of the magnetic moments. This is associated with an increase in energy and thus results in an intermittent relaxation path. Each individual magnetic moment can be represented by a positive and negative charge (indicated as red and blue points on the nanomagnets). For the magnetic configurations shown in Fig. 1(e) and (g), there are two positive and two negative charges at the vertex, and therefore no net charge, whereas the higher energy configuration in Fig. 1(f) is associated with a net charge at the vertex highlighted in red.

The relaxation paths between different states can be depicted using a state network diagram for the energy landscape, as shown in Figs. 1(d) and 1(h) for the 4-loop and the 4-vertex structures, respectively. Here, states are indicated with black points that are located according to their energies $E_{total}$ along the vertical axis and the net magnetization $\pm|M| = |\sum_i^N m_i|$ of the state along the horizontal axis. The sign of $\pm|M|$ indicates whether $M$ is parallel or antiparallel to the initial applied field $H$ (blue arrows in Fig. 1(a,e)). Furthermore, it should be noted that a randomized offset along the horizontal axis was added to every point in order to avoid overlap of degenerate states. Single-moment-reversal connections between states $m$ and $l$ are represented by tapered lines, where the path is travelled from the wide to the narrow end. The center of each line corresponds to the energy barrier $E_m + E_T$, and the grayscale of the line indicates the transition rate $\nu(E_T, T)$ at $T = 295$ K according to Eq. (2), calculated using the parameters specified in the Methods section, and given by the grayscale bar. It should be pointed out that the system encounters energy barriers for all paths regardless of the curvature depicted in the state network diagrams.

Boolean operations can be realized with nanomagnet structures exhibiting a monotonous relaxation pathway, which guides the relaxation towards a specific final state. In contrast, intermittent relaxation pathways lead to a probability of achieving particular low-energy states, which can be implemented in computational schemes where weighted outputs are desired such as artificial neural networks.

It should be noted that the relaxation pathways may incorporate metastable states, indicated by the triangles in the state network diagrams in Figs. 1(d) and (h). During relaxation to the degenerate ground states, the structures can become trapped in intermediate states that are not suitable for use in Boolean computation since, for a given input, the output does not always conform to a Boolean operation.

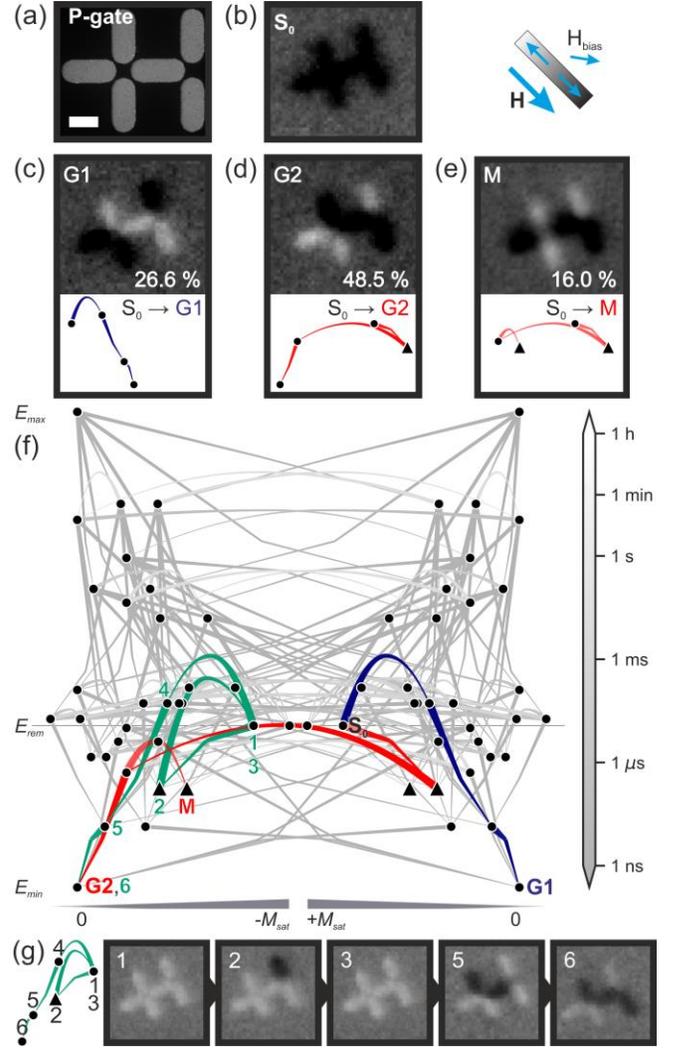

**Fig. 2** Thermal relaxation and state network diagram of the P-gate. (a) Scanning-electron microscopy (SEM) image of the P-gate. The scale bar measures 250 nm. (b) Field-set initial state $S_0$, degenerate ground states (c) G1 and (d) G2, and metastable state (e) M, imaged by X-PEEM. For 390 gates, G1 is reached in 26.6 % of the gates, whereas G2 is reached in 48.5 % of the gates. In addition, 16.0 % of the gates get stuck in the metastable state M. The relaxation path taken is determined by the height of the successive activation barriers that need to be overcome during relaxation. (f) State network diagram of the P-gate at $T = 295$ K. The fastest relaxation path from the field-set state $S_0$ to the ground states G1 and G2, represented by the larger black points, are intermittent paths highlighted in blue and red. (g) Many different relaxation paths can be explored and there can even be a reversal of direction with the system travelling back along a given path. This is illustrated with successive imaging of the relaxation from an initial field-set state (1) to a final low-energy state (6). Here a change in direction of relaxation was observed when transitioning from (1) to (2) to (3) and, subsequently, the structure traverses the path to G2. The corresponding path in (f) is highlighted in green.

## III. METHODS

Arrays of elongated Permalloy nanomagnets, with a length of 470 nm and a width of 170 nm, were fabricated on a silicon substrate using electron beam lithography in combination with lift-off processing. The nanomagnet arrangements in this work are based on an artificial square ice design, with the nanomagnet center-to-center distance a = 600 nm [see Fig. 1(c)]. In order to experimentally observe the thermal relaxation, in particular the individual moment reorientations at an accessible timescale, the nanomagnets arrays were fabricated from a Permalloy wedge film with a thickness ranging from 1 to 15 nm, which was deposited using thermal evaporation at a base pressure of $2 \cdot 10^{-6}$ mbar and capped with 2 nm Aluminum to prevent oxidization.

Magnetic imaging of the nanomagnets arrays was performed using X-ray photoemission electron microscopy (X-PEEM), which makes use of resonant X-ray magnetic circular dichroism (XMCD) at the Fe $L_3$ edge to obtain magnetic contrast [25]. A pixelwise division of the intensity in the electron yield, for right and left circularly polarized incident X-rays, gives a contrast that corresponds to the orientation of the magnetic moments. For observations of thermal relaxation, a particular film thickness at a given temperature was chosen with moment reorientations at a second to minutes timescale, so that magnetic configurations between moment reorientations could be imaged.

The following experimental protocol was implemented to probe thermal relaxation behavior in different arrangements of nanomagnets. After field-setting the nanomagnets, so that all moments align towards the direction of the applied magnetic field, the sample was heated in order to assist thermal relaxation to a state of lower dipolar energy [17,26]. The final configurations were imaged with X-PEEM for a large number of equivalent arrays.

Experimental observations were compared to theoretical predictions based on kinetic Monte-Carlo simulations [26], using the point-dipole Hamiltonian of Eq. (1), and the model of thermally-assisted moment reorientation described by Eq. (2). For the simulations, a lattice parameter of $a$=600 nm was used (Fig. 1), volume magnetization was set to $M_S = 350 \cdot 10^3$ Am$^{-1}$ (corresponding to ~$9 \cdot 10^6$ $\mu_B$ per nanomagnet moment), the attempt frequency to $\nu_0 = 10^9$ s$^{-1}$, and the single-nanomagnet switching barrier to $E_b = 0.626$ eV [23,26]. The temperature for kinetic Monte Carlo simulations was set to 460 K. For the visualization of state network diagrams, the same values of $a$, $M_S$, $\nu_0$, $E_b$ were used with transition rates calculated for T=295 K. To obtain transition probabilities in agreement with experimental observations, we have to assume that there is a small effective bias magnetic field ($H_{bias}$) of ~ -50 μT along the direction shown in Fig. 2. The presence of such a spurious field in the experimental X-PEEM chamber has been confirmed in recent experiments [27].

## IV. RESULTS AND DISCUSSION

### (a) Probabilistic Operations

We begin by presenting the relaxation dynamics of the design shown in Fig. 2(a), which we refer to as a probabilistic logic gate or P-gate. Starting from a field-set state $S_0$ [Fig. 2(b)], and following thermal relaxation at ~290 °C for 2 hours for a gate with ~5 nm thick nanomagnets, we found that of the 390 P-gates tested, 26.6 % of the gates reached the ground state G1 [Fig. 2(c)], and 48.5 % of the gates ended up in the degenerate ground state G2 [Fig. 2(d)]. In addition, several other states were observed after relaxation, most notably the metastable state M in Fig. 2(e), which accounted for 16 % of the final configurations.

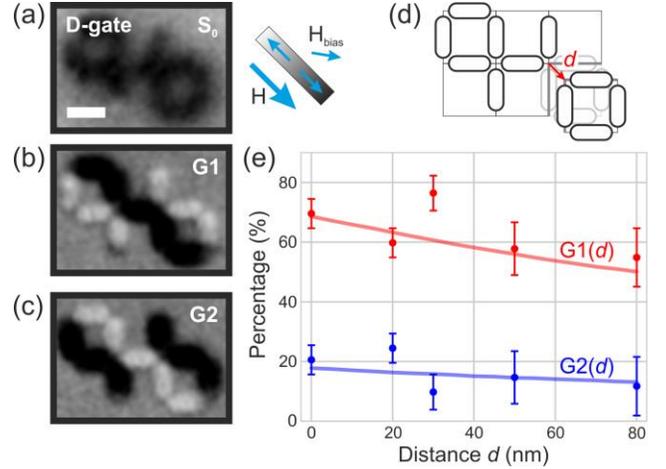

**Fig. 3** Tuning of ground state occupation with modification of the nanomagnetic design. X-PEEM images of (a) the field-set initial state $S_0$ of the original D-gate design, and the ground states (b) G1 and (c) G2. The scale bar in (a) measures 500 nm. (d) Schematic drawing of the D-gate. In order to tune the probability of structures reaching the ground state G1, the lower right 4-loop is moved by the distance $d$. (e) The percentage of gates ending up in the two ground states G1 and G2 for a 4-loop offset by a distance d. The final percentages of low energy states in 51 gates for different distances (points) was compared with the kinetic Monte-Carlo simulations (solid lines).

The lower probability of reaching G1 as compared to G2 indicates that the transition barriers towards G1 must be higher than those that occur on the path towards G2. This is illustrated in the state network diagram in Fig. 2(f). Here, the fastest paths from $S_0$ towards G1 and G2, represented by the larger black points, are highlighted in blue and red respectively. Both paths include three moment reversals and are of the intermittent type. The first transition on the path towards G1 (in blue) requires a higher thermal activation energy than the one towards G2 (in red), thus biasing the system towards G2.

The relaxation dynamics characterizing the exploration of competing relaxation paths by the P-gate is experimentally shown in the image sequence in Fig. 2(g), where individual moment reorientations during relaxation are imaged, and is indicated by the path highlighted in green in Fig. 2(f). Here the P-gate was field set with a magnetic field applied in the opposite direction (180°) to the field applied in Fig. 2(g) [Image 1]. During the course of relaxation, the P-gate initially traverses the path towards G1 but, as it encounters a metastable state, corresponding to the second image in Fig. 2(g), its course is reversed by traveling back to the initial state and then traversing the path to G2.

In summary, intermittent paths in the state network diagram result in probabilistic relaxation. The direction of an applied field is also important since this defines the initial state and thus the subsequent energy landscape. Finally, it should be pointed out that the intermittent relaxation path for the P-gate is a result of the incorporation of the 4-vertex structure shown in Fig. 1(e) into the gate design, which leads to the creation of a charge at a vertex as the gate transitions from an initial field set state to the final low energy state during the course of thermal relaxation.

### (b) Deterministic Operations

Deterministic relaxation towards a specific ground state can be realized by introducing the 4-loop structure shown in Fig. 1(a) into the design to give a monotonous relaxation path. To demonstrate this idea, we use the structure shown in Fig. 3(a), which we refer to as a deterministic logic gate or D-gate. With the D-gate, we previously demonstrated Boolean operations, with more than 90% of D-gates experimentally observed to relax to the G1 state [17]. The high operational reliability of the D-gate results from that fact that the first relaxation step from the initial field-set state $S_0$, [Fig. 3(a)], towards the low energy state G1 [Fig. 3(b)], has a lower energy barrier than the first transition on the relaxation path towards the low energy state G2 [Fig. 3(c)]. Therefore, although the path towards G1 requires six moment reversals, compared with the five reversals required to reach G2, the system is kinetically biased to relax towards G1 [see supplementary data in Ref. 18].

In summary, structures with monotonous paths in the state network diagram result in deterministic relaxation. Even though the D-gate has two vertex structures, any vertex charge creation in the two vertex structures does not lead to an increase in energy in the D-gate, as the change in energy during charge creation is counter-balanced by the lowering of the dipolar energy in two 4-loop structures.

### (c) Continuous Modulation of Output

The relaxation outcome of the D-gate can be modified, by physically altering its design. This can be achieved through a change in the position of one of the two 4-loop structures so that it is at different distances $d$ away from the vertex point [see Fig. 3(d)]. Increasing $d$ results in a weakening of the dipolar coupling between the 4-loop containing the output and rest of the gate, leading to a higher probability of accessing non-ground-state low-energy states along the path towards G1, and thereby reducing the final percentages of low-energy states, $p_{G1}$.

We tested the thermal relaxation behavior of 51 modified D-gates for each $d$, for different separations $d$ between 0 nm and 80 nm. It should be noted that the thermal protocol was different than that used in our earlier experiments [17]. Due to experimental constraints, a lower temperature of $T\sim190$ °C was used along with a thickness of ~3 nm for the nanomagnets and a relaxation time of an hour, as opposed $T\sim290$ °C, nanomagnets thickness ~5 nm and a relaxation time of two hours. The percentages $p_{G1}$ and $p_{G2}$ of ground states G1 and G2, as a function of separation $d$, are plotted as points in Fig. 3(e). We observe a decrease of $p_{G1}(d)$ with increasing $d$ from ~70 % at 0 nm to ~55 % at 80 nm, and less variation with $d$ for $p_{G2}(d)$. These percentages for the modified D-gates were compared with kinetic Monte-Carlo simulations [solid lines in Fig. 3(e)], which reproduce the experimental results well with the available statistics using the parameters given in the methods section. This demonstrates that the target probabilities can be controlled by physically modifying the gate design.

### (d) Extended Circuits

So far, we discussed the relaxation behavior of individual computational building blocks containing a few nanomagnets each. To create extended circuits capable of advanced calculations, such as adders or artificial neural networks, these individual gates need to be combined into larger structures.

In order to integrate more than one gate into a larger circuit, several conditions need to be met. First, each individual logic gate requires two or more input magnets and an output magnet. Second, the input and output magnets of different gates need to be at a location that allows them to be linked to neighboring gates. Third, there needs to be a strong coupling between the output magnet of one gate and the input magnet of the next gate. Finally, their respective relaxation kinetics, which governs their operational reliability, should remain largely unaffected when the gates are linked together.

An experimental realization of three D-gates connected by linear chains of strongly-coupled nanomagnets is shown in Fig. 4. Starting from the field-set state, shown in Fig. 4(b), the extended circuit relaxes into a ground state. This result provides a first proof-of-concept for integrating logic gates into extended arrays for use in large circuits.

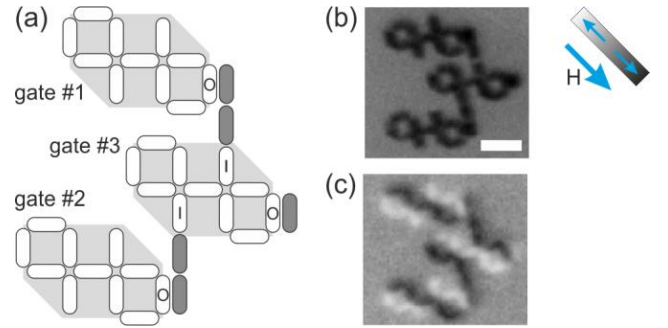

**Fig. 4** Connecting several D-gates. (a) Three equivalent D-gates (highlighted in light gray) are coupled by strongly-interacting parallel nanomagnets (in dark gray). These link the output magnets (O) of the outer two gates #1 and #2 to the input magnets (I) of the middle gate #3. (b) Initial field-set state, and (c) final state imaged with X-PEEM. After thermal relaxation, the outer gates #1 and #2 relax to the ground state G1, and the central gate #3 relaxes to G2. Scale bar in (b) measures 1 μm.

## V. CONCLUSIONS

In this work, we have shown how different relaxation pathways can be engineered in artificial spin ice structures comprising arrays of dipolar-coupled nanomagnets for deterministic and probabilistic computation. We have identified two distinct categories of relaxation pathways, which can be (1) monotonous and (2) intermittent and were visualized using a nanomagnetic state network diagram.

The implementation of loop-based designs in the D-gates, which favors monotonous relaxation pathways, greatly improves the reliability of Boolean logic gates as demonstrated previously [17]. In contrast, intermittent relaxation pathways such as P-gates, often associated with the creation of energetically unfavorable vertex charges, lead to different probabilities of low energy states, as the system can explore competing relaxation pathways. The different probabilities to reach particular low energy states could be exploited in weighted computational schemes such as artificial neural networks.

We have also demonstrated that the probability of low-energy states can be tuned by changing the logic gate design. This also depends on, for example, the direction of the applied magnetic field used to set the initial state of different gates. This control could also be achieved in-situ, for example with Oersted fields in a current-carrying nanowire [28] or exploiting strain-mediated effects [9]. Such schemes would even provide a means to vary the transition barriers between or even during computational (i.e. relaxation) cycles.

Finally, the integration of individual logic gates into extended circuits opens the possibility for implementation in both conventional computation [4] as well as in novel computational schemes such as those involving mapping a problem to a magnetic Hamiltonian [13,29].

## VI. ACKNOWLEDGEMENTS

This work was supported by the Swiss National Science Foundation. XPEEM experiments were performed at the Surface/Interface: Microscopy (SIM) beamline of the Swiss Light Source, Paul Scherrer Institute, Villigen, Switzerland. H.A. and N.L. are supported by SNSF Grants 200021_155917 and 200020_172774. D.S. has received partial funding via a PSI-CROSS proposal (Grant No. 03.15). J.V. is supported by SNSF Grant 200021_153540. J.C. has received funding from the European Union's Horizon 2020 research and innovation program under the Marie Sklodowska-Curie grant agreement No. 701647.